\documentstyle[prd,aps,psfig]{revtex}
\bibstyle{unsrt}

\tighten
\begin{document}
\draft

\preprint{LAUR-99-6369}
\twocolumn[\hsize\textwidth\columnwidth\hsize\csname 
@twocolumnfalse\endcsname

\title{The Ginzburg regime and its effects on topological defect formation}
\author{Lu\'{\i}s M. A.  Bettencourt$^1$, Nuno D. Antunes$^2$, 
and  W.H. Zurek$^1$}
\address{$^1$T-6, Theoretical Division, Los Alamos National Laboratory,
Los Alamos NM 87545,USA}
\address{$^2$Blackett Laboratory, Imperial College, London SW7 2BZ, 
United Kingdom. }
\date{\today}
\maketitle

\begin{abstract}

The Ginzburg temperature has historically been proposed 
as the energy scale of formation of topological defects at a second order
symmetry breaking phase transition.
More recently alternative proposals which compute the time of
formation of defects from the critical dynamics of the system \cite{Zurek}, 
have been gaining both theoretical and experimental support. 
We investigate, using a canonical model for string formation,
how these two pictures compare. In particular we show that prolonged exposure
of a critical field configuration to the Ginzburg regime results in no
substantial suppression of the final density of defects formed. 
These results dismiss the recently proposed role of the Ginzburg regime in 
explaining the absence of topological defects in  $^4He$ pressure quench
experiments. 
\end{abstract}

\pacs{PACS Numbers : 05.70.Fh, 11.27.+d, 98.80.Cq \hfill   LAUR-99-6369}

\vskip2pc]

\section{Introduction}
\label{intro}

The Ginzburg temperature $T_G$ \cite{Ginzburg} 
was thought historically to be the
determining energy scale at which topological defects are formed in the 
aftermath of a second order symmetry breaking phase transition 
\cite{K76,K80,Book}.
More recently theoretical and experimental progress has pointed in the 
direction that it is the critical dynamics of the second order transition, 
induced by a change in some external parameter such as 
temperature or pressure, 
that determines the numbers (and configuration) of topological defects 
emerging below the critical point \cite{Zurek}.
 
Nevertheless the role of large thermal fluctuations within the Ginzburg 
regime in defect formation mechanisms remains relatively poorly understood. 
In particular it is not clear how a density of defects created, presumably
by the critical dynamics of the system, could evade alteration when exposed
extensively to thermal fluctuations in the Ginzburg regime.

This issue has been rekindled recently due to the possibility of 
new experimental tests \cite{He3G,He3H,He4old,He4new} and, in particular, 
by the negative results of a pressure quench experiment 
in $^4He$ \cite{He4new}, a system in which
(because of strong interactions) the Ginzburg regime is particularly
extensive.
This experiment has improved on an apparatus used earlier by 
McClintock {\it et al}. \cite{He4old} to implement a superfluid
transition in $^4He$ through a sudden pressure quench. 
The results show no evidence for the 
formation of topological defects at the anticipated levels,
contrary to expectations based both on the old experiment 
\cite{He4old}, the theory\footnote{Although a factor 
$f\stackrel{>}{\sim} 10$ in the formula for the string 
density $n\sim 1/(f \hat \xi)^2$ could explain the new results and  
seems consistent with recent numerical studies \cite{ABZ}.} 
and the $^3He$ data \cite{He3G,He3H}.

The discrepancy with the earlier $^4He$ quench data is now seen as the 
evidence of mechanical stirring in the first version of the experiment.
Nevertheless to address this discrepancy with $^3He$ it was 
suggested \cite{KR} that because the Ginzburg regime in $^4He$ 
extends over a broad range of temperatures around the $\lambda$-line, 
large scale fluctuations may be able to unwind  and alter the configuration 
of the order parameter (in contrast to $^3He$) while  the quench proceeds.
The Ginzburg temperature is defined, somewhat qualitatively, 
through the loss of ability of the order parameter to hop, 
through thermal activation, over the potential barrier between 
broken symmetry minima. Thus one might worry with Karra and Rivers \cite{KR} 
that when the defect densities are eventually measured, at a much later time, 
little or no string would have survived unwinding through thermal activation.

In in this paper we study in detail the role of the Ginzburg regime 
in vortex string formation. In section \ref{FP} we discuss our model
and its properties. We show in particular that it transcends the more usual
time dependent Ginzburg-Landau (TDGL) dynamics in generality and 
reduces to it  in particular cases.
In section \ref{thermo} we describe 
the traditional arguments for the existence of a well defined 
Ginzburg temperature and critically analyze their underlying  
assumptions in the light of known results on the thermodynamics 
of vortex strings. We also establish  a quantitative definition of 
the Ginzburg temperature and discuss its uncertainties. In section 
\ref{dynamics} we investigate the role of the Ginzburg regime in the
{\it dynamics} of defect formation. This is achieved by exposing field 
configurations created at criticality to intermediate temperatures
within the Ginzburg regime and analyze the effect upon the final
density of defects measured at late times. 
We also study the memory of the order parameter when subjected to
reheating. This constitutes a direct test on the theory of 
defect formation as a consequence of the critical dynamics of the theory.
Finally we draw our conclusions and discuss in the light of our
results the possible relevance of the Ginzburg regime in explaining  
recent experimental results in $^4He$ pressure quench experiments.

\section{Langevin and Fokker-Planck Field dynamics}
\label{FP}

As a working model we consider a $U(1)$ symmetric $\lambda \phi^4$ global
field theory in 3 spatial dimensions (3D), in the canonical ensemble, 
i.e. in contact with a heat bath at a given temperature $T$. 
This model has the advantage 
of having been extensively studied in thermal equilibrium 
\cite{ABH,AB,XY,Tsubo} and moreover of describing the 
thermodynamics of $^4He$ at criticality \cite{Zinn} by 
permitting the computation of relevant critical exponents with extreme
accuracy.

As such it supplies us with a controlled realistic environment in which 
the role of thermal fluctuations within the Ginzburg regime in changing 
the density of topological defects can be studied.
The equations of motion for the fields are 
\begin{eqnarray}
&&\left[ \partial_t^2  + \eta \partial_t  -\nabla^2 - m^2 \right]
\phi_i(x)    \nonumber \\
&& \qquad \qquad + \lambda  \left(\sum_{j=1}^2 \phi_j^2(x)
-1 \right) \phi_i(x) = \xi_i(x,t), \nonumber \\ 
&& \langle \xi_i (x,t)\rangle = 0, \nonumber \\
&& \langle
\xi_i(x,t) \xi_j(x',t')  \rangle = \Omega \delta(x-x') \delta (t-t')
\delta_{ij}.
\label{eq1}
\end{eqnarray}
where $i,j \in \{1,2\}$ and the heat bath fields $\xi_i(x,t)$ 
obey the fluctuation dissipation relation
in equilibrium.  Thus, for long times, the system equilibrates to its
canonical distribution at temperature $T$. This can be shown most
conveniently by writing the Fokker-Planck equation corresponding to
the Langevin dynamics of Eq.~(\ref{eq1}) \cite{Bett},
\begin{eqnarray}  
\partial_t P_{FP}[\pi,\phi,t]=-{\cal H}_{\rm FP} P_{FP}[\pi,\phi,t].
\label{e10}
\end{eqnarray}
where
\begin{eqnarray}
{\cal H}_{\rm FP} &=& - {\Omega \over 2} 
{\delta^2 \over \delta \pi_i^2 } + \pi_i {\delta \over \delta 
\phi_i} \nonumber \\
&& \qquad \qquad - {\delta \over \delta \pi_i} 
\left( \eta \pi_i - \nabla^2 \phi_i + 
{\delta V(\phi) \over \delta \phi_i}\right), 
\label{e11}
\end{eqnarray}
where sum over $i\in \{1,2\}$ is implied here and below.
If, as in most applications, the potential $V(\phi)$ is explicitly time 
independent we can invoke a separation ansatz for $P_{FP}$ such that 
\begin{eqnarray}
P_{FP} [\pi,\phi,t] = {\cal P} [\pi,\phi] T(t)
\label{e12}
\end{eqnarray}   
Thus we  can regard Eq.~(\ref{e10}) as a functional Schr\"odinger 
equation, in imaginary time. 
Then we can write the time independent and dependent equations 
\begin{eqnarray}
{\cal H}_{\rm FP} {\cal P}_n = E_n {\cal P}_n, \quad \partial_t T(t) = 
- E_n T(t).
\label{e13}
\end{eqnarray}
The functional dependence on the fields is now limited to 
the static probability eigenfunctionals ${\cal P}_n$. The time evolution
of the Fokker-Planck distribution is completely characterized by the spectrum
of eigenvalues of ${\cal H}_{\rm FP}$, $E_n$.  

Formally, we can then project the evolution of $P_{FP}$ in terms of its 
eigenvalues $E_n$ and eigenfunctionals ${\cal P}_n$ as:
\begin{eqnarray}
P_{FP}[\pi,\phi,t] = \sum_{n=0} ^\infty C_n  {\cal P}_n[\pi,\phi]  
e^{- E_n t}.
\label{e14}
\end{eqnarray}
where the $C_i$'s are the projections of the initial time $P_{FP}$ onto
the basis of eigenfunctionals ${\cal P}_n$.

The equilibrium solution must be static. It corresponds to a 
zero eigenvalue in Eq.~(\ref{e13}), which implies the solution 
\begin{eqnarray}
P_{\rm eq} [\pi,\phi] = N \exp \left[ -\beta \int d^D x {\pi_i^2 \over 2} + { 
(\nabla \phi_i)^2 \over 2} +  V[\phi] \right],  
\label{e17}
\end{eqnarray}
where we took $\Omega=2 \eta/\beta$, which is the Einstein relation
enforcing equilibrium between fluctuation and dissipation at long times.
Summation over $i$ is implied.
On general grounds we expect the canonical equilibrium distribution
to be approached at long times, i.e. we expect the excited time-dependent 
states $P_n$, $n \neq 0$ to decay with time.

The full spectrum of excited states and their corresponding eigenvalues
can only be found for specific forms of the field potential $V(\phi)$.
In particular this is possible in closed analytic form for harmonic 
potentials $V= {1 \over 2} m^2 \phi^2$. 
For each mode in k-space the excited states are given in terms of 
Hermite polynomials of functions of the field modes and those  of 
their conjugate momenta. More importantly the corresponding 
eigenvalues are given by \cite{Bett}
\begin{eqnarray}
E_n &&= n {\eta \over 2} \left[1\pm \sqrt{1 
-4 (k^2 +m^2)/\eta^2}\right].
\nonumber 
\end{eqnarray}

Close to the phase transition the leading effect of the 
${\lambda \over 4} \phi^4$ interactions is to make the 
effective mass temperature dependent as 
\begin{eqnarray}
m^2(T) = m_0^2 \vert {T -T_c \over T_c}\vert^{2\nu} 
\label{mass}
\end{eqnarray}
where $T_c$ is the critical temperature and $\nu$  a universal 
critical exponent, which depends only on the dimensionality of 
space and the internal symmetries of the theory.  To 1-loop in perturbation
theory we have
\begin{eqnarray}
m^2(T) = -m^2 + \Delta m^2(T), \nonumber \\
\Delta m^2(T) = \lambda  \int {d^D k \over (2 \pi)^D} \langle \phi_k \phi_{-k} \rangle. 
\label{tadpole}
\end{eqnarray}
i.e. the temperature correction to $m^2$ is given by the (classical)
tadpole diagram.
The values of the $O(2)$ symmetric thermal average 
$\langle \phi_k \phi_{-k} \rangle$ depends on the specific
form of the thermal distribution, classical or quantum. 
Under these approximations one obtains the mean-field value of $\nu=1/2$.

In the critical domain we can thus obtain an approximate analytical 
description of the non-linear field dynamics by taking the mass in the 
harmonic potential to be of the form (\ref{mass}). 
Although only approximately 
true we will show below that this assumption leads to a good qualitative 
understanding of the full non-linear field dynamics in the critical domain 
and the effects of  the Ginzburg regime.

The present scheme, Eqs.~(\ref{eq1}), 
is therefore convenient both as a thermalization
algorithm, if the  system is started at any given configuration and
run for long times, or as a means of performing  effective
non-equilibrium dynamics. The equilibration time itself $t_{eq} \simeq 1/E_1$,
is dependent  on spatial scale (or wave length) and on temperature. 
Qualitatively large spatial scales equilibrate more slowly and in 
particular, in the critical domain, exhibit critical slowing down.
This can be seen explicitly by considering long wave-length modes
($k^2 \simeq 0$) in the vicinity of $T_c$. Then, for half of the
eigenvalues the termalization time is inversely proportional to $n$ times
$t_{eq}$ with 
\begin{eqnarray}
t_{eq} = && 
\left[ {\eta \over 2} \left(1-\sqrt{ 1 - 4 (k^2 +m^2)/\eta^2} 
\right) \right]^{-1} \\
&& \simeq {\eta \over m^2(T)} \rightarrow \infty,
\end{eqnarray}  
as $T\rightarrow T_c$.
This is the result for overdamped dynamics $\eta >> m(T)$, and could have 
been obtained by a Langevin equation with a single (dissipative) 
time derivative. In this sense the evolution of the long-wave length modes
in the vicinity of $T_c$ is always overdamped, which is the essence of the 
perhaps more familiar TDGL evolution, to which Eqs.~(\ref{eq1}) reduce to 
in the appropriate regime. Note that the TDGL dynamics is an effective
equation for long-wave length field modes in the critical domain while our 
model holds more generally.

In the converse limit the decay of short 
wave-length modes is dictated by $\eta$ and is thus scale invariant 
in this approximation. The appropriate physical value of $\eta$ can be 
computed in perturbation theory given a microscopic model. 
Particularly interesting are 
situations for which $m(T) < \eta$ as happens in the critical domain. Then 
there is true time-scale separation in the sense that short wavelength modes
thermalize much faster than long-wave length modes. 

This kind of considerations will help us understand the behavior 
of the fully non-linear dynamics in the Ginzburg regime. Before  
we do this we need to develop a clear picture of equilibrium 
to which we now turn.

\section{Equilibrium results and the definition the Ginzburg temperature}
\label{thermo}

The rationale behind the original proposal according to which the Ginzburg 
temperature $T_G$ is the  energy scale for the formation of 
topological defects \cite{K76,K80},
was that, at lower temperatures, thermal fluctuations would be unable 
to overcome the potential energy barrier associated with the defect's
topological stability. Thus, field configurations with non-trivial
topology, below this temperature would necessarily acquire 
stability on the average.  

It is clear that such an appealingly simple physical picture 
assumes implicitly 
a separation of physical scales and associated time evolution 
or equivalently, as we discuss below,  that at least part of the system
is out of thermal equilibrium.
Indeed one must assume that field configurations can be separated in two
populations - one of topological defects and another of thermal fluctuations.
The former, at least in the sense of the definition of the 
Ginzburg regime (see below) live on a zero temperature background.  
This situation is at best an idealization.

\begin{figure}
\centerline{\psfig{file=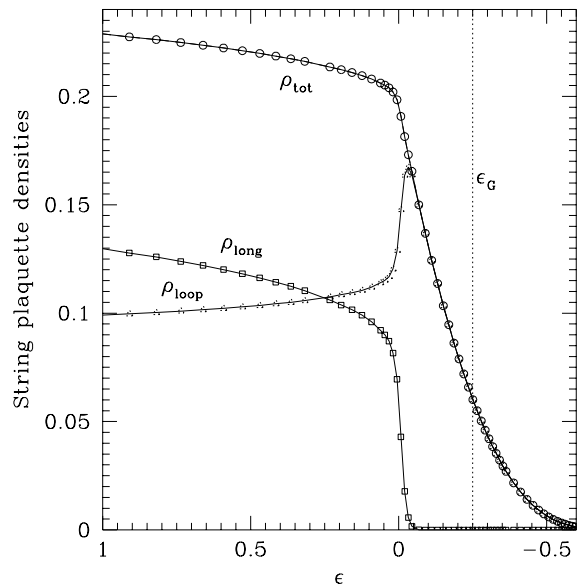,width=3.25in}}
\caption{The dependence of the thermal distribution of vortex string field
fluctuations on temperature ($\epsilon={T-T_c \over T_c}$), 
around the critical point $\epsilon=0$. 
The critical point is marked by the sudden appearance of long strings. 
The observed fast transient should become a discontinuous jump in the 
infinite volume limit [16].}
\label{fig1}
\end{figure}

Strictly in thermal equilibrium 
at temperatures not too low, field thermal fluctuations with 
non-trivial topology will always exist. The density of vortex 
string thermal fluctuations
in our model is shown in Fig.~\ref{fig1}. It is, however, remarkable
that long strings can only exist in equilibrium strictly 
above $T_c$ \cite{ABH,AB}. This
phenomenon is the analog of vortex pair unbinding 
in the well known Kosterlitz-Thouless transition
in this very same model in 2D \cite{KT,Williams}. 
In 3D, however, long strings appear abruptly as we are
dealing with a true critical phenomenon instead of a crossover.

\begin{figure}
\centerline{\psfig{file=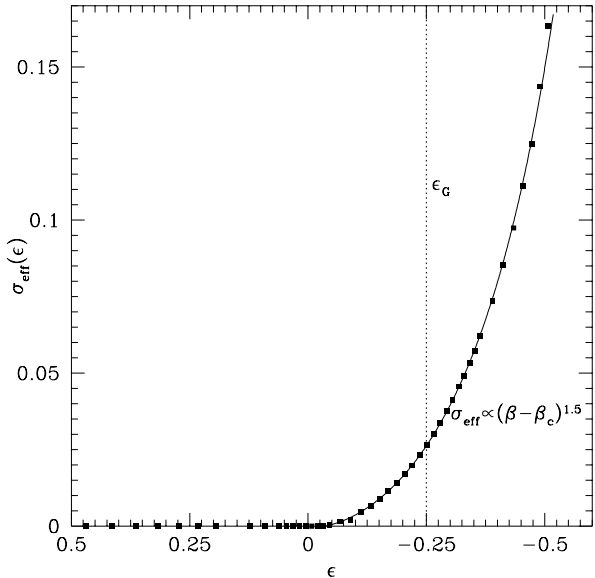,width=3.25in}}
\caption{The dependence of the string tension $\sigma$ 
(free energy per unit length)
on temperature in the critical domain. The critical point is marked
by the vanishing of $\sigma$, which in turn allows for the creation 
of arbitrarily long strings as thermal fluctuations, see Fig.~\ref{fig1}.}
\label{fig2}
\end{figure}

The appearance of long strings exactly at $T_c$ can be understood, in
turn, in terms of their thermal statistical properties namely their 
tension $\sigma_{\rm eff} (T)$ 
(free energy per unit length) and other statistical properties like 
their fractal or Hausdorf dimension \cite{AB}. 
The dependence of the string tension on temperature is shown 
in Fig.~\ref{fig2}. As seen the string tension diminishes continuously  
with increasing temperature until the critical point where it vanishes.  
This permits the creation of strings of all lengths above $T_c$. Below
$T_c$, on the other hand strings are exponentially suppressed and 
only those smaller than the temperature dependent length $l\simeq
T/\sigma_{\rm eff}(T)$ are likely as thermal fluctuations. It is the existence 
of long strings as thermal fluctuations that will lead to defect
formation if the system is suddenly cooled \cite{ABZ}. 

Although some of the above comments may seem somewhat marginal they
establish that the thermodynamics of the $U(1)$ theory under
consideration is much richer than the assumptions on which the
traditional role of the Ginzburg regime is based. They show in
particular that the vortex strings themselves as a subset of the
theory's thermal fluctuations have a very non-trivial thermodynamics
and cannot be taken as their cold classical solutions over a nontrivial 
thermal background.

The thermodynamics of vortex strings in more complex field theories, 
with gauge fields and larger symmetry groups, remains to date 
largely unstudied, although some work has been done 
in the Abelian case \cite{Fin}. 
We expect nevertheless that most of the features 
of our  $U(1)$ global theory may persist albeit characterized by
different critical exponents. This statement is supported by analytic 
studies of the statistics of {\it free} strings \cite{Ray,Turok}, 
which, thanks to their large configurational entropy, exhibit a similar 
(Hagedorn) transition but display eg. a different dependence of the
string tension on temperature. The interactions thus change particular
temperature dependences of certain quantities but not their 
qualitative behavior. In theories where 
defects are not topologically stable however (as in the case of
embedded or semilocal defects) the role of these configurations may be 
potentially different and possibly marginal.

The cumulative results from the equilibrium study of the
thermodynamics of vortex strings in our model cast considerable 
doubt upon the assumptions underlying the 
traditional role of the Ginzburg temperature in defect formation. 
It remains unclear however what the role may be of large thermal 
fluctuations in the critical domain (above $T_G$) in changing defect
densities produced by the critical dynamics of the fields. 

In order to investigate this issue we need a quantitative 
definition of $T_G$. In tune with the arguments given above 
consider a volume of characteristic size $\xi (T)$, the correlation length,  
and a theory with two energetically degenerate minima of an 
effective potential $V(\phi)$, separated by a potential barrier $\Delta V$.
This applies also for theories with a general $O(N)$ symmetry, since we
will be interested in the 'radial' direction only. The effective 
potential is obtained by singling out an arbitrary direction in field 
space \cite{EW}, which we denote by $\varphi$. 
The rate for the field to change coherently from one minimum 
to the other per unit volume due to thermal activation is 
\begin{eqnarray}
T^4 \exp \left(- \Delta V/k_B T  \right).
\end{eqnarray}
For an effective potential of the form (obtained, eg. perturbatively at 1-loop)
\begin{eqnarray} 
V(\phi) = -{1\over 2} m^2(T) \varphi^2  + {\lambda  \over 4} \varphi^4  
\label{Veff}
\end{eqnarray} 
$\Delta V = {m(T)^4 \over 4 \lambda}$. For a volume $\xi^3$, we define 
$T_G$ such that the probability of overcoming the potential barrier 
is of order unity:
\begin{eqnarray}
T_G \ : \  {\Delta V(T_G) \over T_G} . \xi^3(T_G) =1 \  
\Leftrightarrow  \ \lambda T_G/m(T_G)={1\over 4}.
\label{Gzg}
\end{eqnarray} 
This definition however has some caveats, for instance, an 
effective potential of the form Eq.~(\ref{Veff}) is only valid for 
the mean field and not on smaller scales. 
A more careful accounting of scales leads to different 
results \cite{Bettencourt}, which show an enhancement of the hoping 
probability. Thus, the factor of $1/4$ in Eq.~(\ref{Gzg}) should 
not be taken at face value.

A perhaps more rigorous definition arises from the range 
of temperatures below $T_c$ 
for which fluctuations are large and consequently where perturbative 
finite temperature field theory fails to be useful.
In order to set up a perturbative scheme at finite temperature from an initial
3+1 dimensional quantum field theory one implements dimensional reduction
which is valid provided the temperature is high compared to all mass scales.
As a consequence the coupling of the dimensionally reduced 3D field 
theory becomes dimensionful, i.e. $\lambda \rightarrow \lambda T= \lambda_3$.
In order to proceed one has to identify an appropriate 
dimensionless coupling. This is done by taking $\lambda T/m(T)$.
The Ginzburg regime is entered when this 3D effective coupling becomes strong,
in the vicinity of the critical point, namely 
\begin{eqnarray}
T_G \ : \ \lambda T_G/m(T_G)=1.
\label{Gz}
\end{eqnarray}
To compute $T_G$ one needs the scaling of $m(T)$ in the 
critical domain. We write 
$m^2(T) = m_0^2 \vert \epsilon \vert^\nu$, with 
$\epsilon$ being the reduced 
temperature $\epsilon= {T-T_c \over Tc}$.

Thus $\epsilon_G= -0.18$  for $\nu=0.5$.
This mean-field estimate produces  an upper bound in $T$ 
for $T_G$ (and lower bound for $\beta=1/T$). 
For realistic 3D exponents, $\nu=0.67$, we obtain $\epsilon_G=-0.25$. 
The first criterion, based on the hopping of a correlation sized volume, 
results in higher values of $T_G$. This brings about a relatively
large uncertainty in the value of $T_G$, which is $18-25 \%$ below $T_c$.

\section{The role of the Ginzburg regime in the dynamics of defect formation}
\label{dynamics}

In order to investigate the role of the Ginzburg temperature
in the {\it dynamics} of defect formation we perform a series of 
tests both directly over the evolution of string densities and   
the evolution of the order parameter, when exposed to thermal
fluctuations in the Ginzburg regime. 

\subsection{Strings Survive the Ginzburg Regime}

To investigate the effects of thermal fluctuations directly 
upon strings we deliberately expose 
the system to a heat bath at temperature $\epsilon_i$, within the Ginzburg 
regime and below. 

\begin{figure}
\centerline{\psfig{file=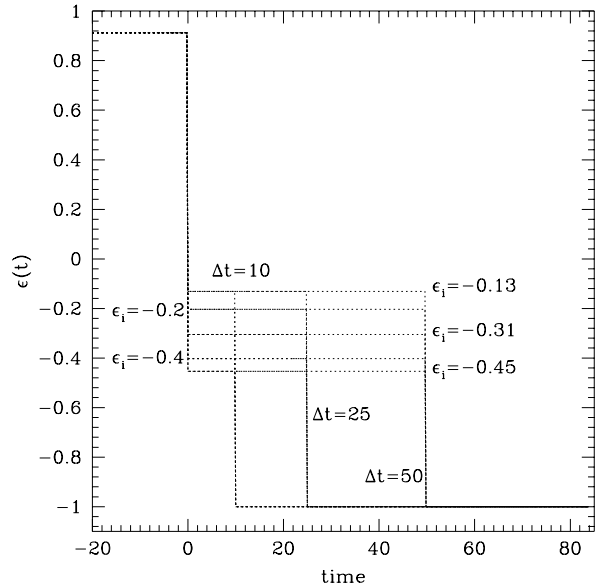,width=3.25in}}
\caption{Temperature trajectories for testing the effect of 
exposure to the Ginzburg regime on string densities. The system is 
first thermalized at a high temperature and then placed in contact 
with a heat bath at an intermediate temperature $\epsilon_i$ below $T_c$,
for a time interval $\Delta t$.}
\label{fig3}
\end{figure}

We are attempting to emulate the worst case scenario of an 
experimental quench where the temperature or pressure are dropped 
monotonically but where the system makes a long stopover within the 
Ginzburg regime. 
We repeat this procedure for a range of time 
intervals $\Delta t$, after which the bath temperature is taken to zero.  
This set of temperature trajectories is shown in Fig.~\ref{fig3}.

We would expect that, if the Ginzburg regime indeed produced  
enhanced decay of strings, then the string densities measured at 
later times should be smaller the longer the time the system spent 
within the range $T_c \geq T\geq T_G$.

\begin{figure}
\centerline{\psfig{file=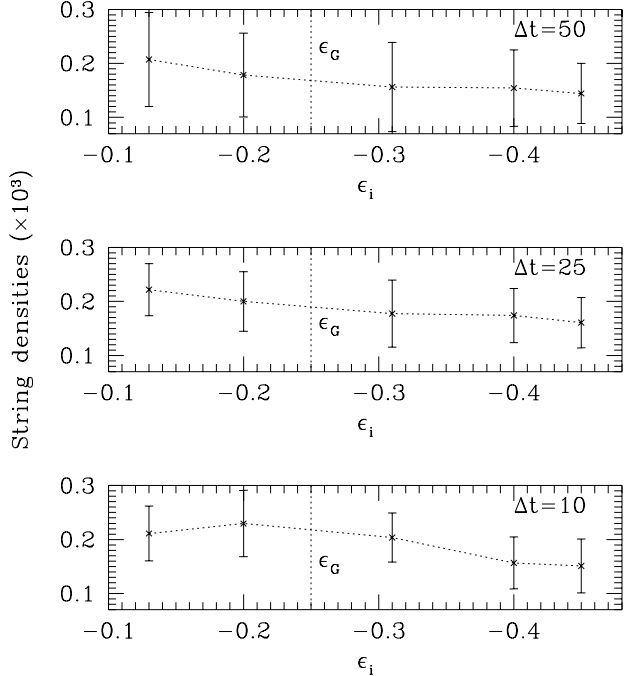,width=3.75in}}
\caption{The string density measured at the same later time $t >> \Delta t$
vs. intermediate temperature $\epsilon_i$. From top to bottom the three 
plots correspond to $\Delta t=10,20,50$, during which the system remained
in contact with a heat bath at $T_i$. There is no visible 
role played by intermediate temperatures within the Ginzburg regime.}
\label{fig4}
\end{figure}

Our results for the final string densities as a function
of intermediate temperature $\epsilon_i$ 
and $\Delta t$ are summarized in  Fig.~\ref{fig4}.
There is no apparent effect of the Ginzburg regime in reducing 
string densities at formation. 

If any trend is visible from Fig.~\ref{fig4} it is the opposite, 
namely a monotonic dependence of the final string densities on
$\epsilon_i$ - the lower $\epsilon_i$, the less string is measured at
later times. 

The knowledge of the vortex string thermodynamics and of the 
time response of the fields in the critical domain again helps us
understand this result. Strings and in particular long strings 
are inherited from high temperature (higher than $T_c$) 
topological fluctuations \cite{ABZ}. 

We can now use our knowledge of the Fokker-Planck solution to understand 
the observations of Fig.~\ref{fig4}.
As we discussed above the small scales in the system equilibrate faster
on a characteristic timescale $t \sim \eta^{-1}$. Small scale fluctuations
dominate the thermal average in (\ref{tadpole}), 
which then allows us to take the effective value of $m^2 \simeq  m^2(T_i)$. 

On the other hand, upon cooling through the critical point the large scales
in the system display critical slowing down. This includes in particular the 
long strings in the sample which will be responsible for the signal 
at the time of measurement later.
This slowing down leads to an imbalance 
in the string population out of equilibrium relative to their equilibrium 
counterpart, given by the existence of many more long strings.

This constitutes an excited state (described by ${\cal P}_{n\neq 0}$) 
relative to the true equilibrium of the 
system  at intermediate temperatures below $T_c$. These states will then
decay on a timescale $t_{eq} = E_n^{-1} \sim  \eta/m^2(T_i)$. 
The value of $m^2(T_i)$ is smaller the closer $T_i$ is to $T_c$ and thus 
leads to a longer time scale for the decay of long strings. 

We can then predict a monotonic behavior for the string densities as  
observed in Fig.~\ref{fig4}. 
At $T_G$ in particular $m^2=\lambda T_G^2$, by definition and $t_{eq} \sim 
\lambda T_G^2/\eta$.

Thus the conclusion is inescapable: The longer the time the system 
spends further from $T_c$ the less string it will display at later times
where formation rates are measured.

\subsection{Memory of the Order Parameter Configuration near $T_c$}

An independent test on the possible role of thermal fluctuations 
in affecting string densities consists in reheating 
a quenched system to a temperature around $T_c$ (both below and 
above) and cooling it again at the same rate. 
This process tests the memory of the order parameter as well as that 
of other related quantities (see also \cite{YZ98}), such as defects.

The importance of this test is directly related to the canonical
theory of defect formation as due to the critical dynamics of the
fields. The final density of strings formed at the transition 
are then infered indirectly through the correlation length associated 
with the two-point correlator of the fields.

\begin{figure}
\centerline{\psfig{file=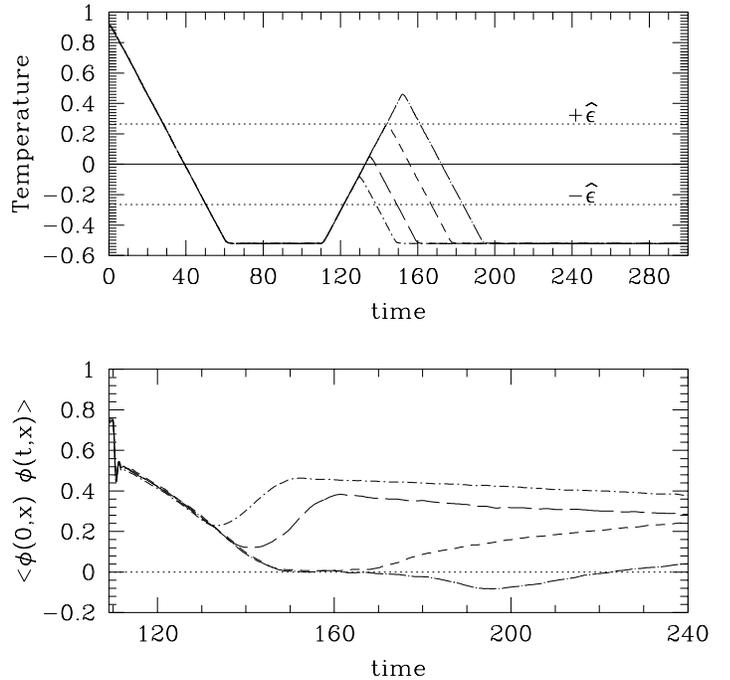,width=3.75in}}
\caption{a. Dependence of the bath temperature $\epsilon$ in 
time. After being quenched in temperature ($\tau_Q=80$)
the system is reheated at the same rate to a temperature 
$\epsilon_f=0.469,0.256,0.061,-0.068$ (top to bottom) and cooled again. 
b. The correlation function between the field at 
the time just before reheating and at later times, 
$\langle \phi_i(t_{\rm rh},x) \phi_i(t+t_{\rm rh},x)\rangle$ is 
plotted. There is a universal short time transient for the decorrelation
of the field over small scales while the long time tails of the
correlation function describe change over the mean fields. 
All four trajectories cross the Ginzburg 
regime, but only those reaching or crossing $+{\hat \epsilon}$ display 
a significant memory loss.}
\label{fig5}
\end{figure}

An example of the temperature ($\epsilon(t)$) trajectories used in
testing the memory of the order parameter are shown in Fig.~\ref{fig5}a.

We are particularly interested in investigating under what circumstances 
thermal fluctuations can affect the large scale configuration 
of the order parameter. 
In order to produce a quantitative test we define the unequal-time 
two-point correlation function
\begin{eqnarray}
&&\langle \phi(x,t_{\rm rh}) \phi (x,t+t_{\rm rh}) \rangle 
\nonumber \\  && \qquad 
=N \sum_{j=1}^2 \sum_i  \phi_j(x_i,t_{\rm rh}) \phi_j(x_i,t+t_{\rm rh}),
\label{cor}
\end{eqnarray}
where $N$ is an irrelevant normalization factor.
This correlator has several interesting properties. 
For short times it displays a characteristic
time, which describes the decay of correlations over very small
spatial scales. This is the initial transient in Fig.~\ref{fig5}b.
We verified that this time and the form of the correlation function 
is in agreement with the forms predicted from a Boltzmann distribution 
for the fields.
For later times the residual correlation comes from the motion
of the order parameter (the field volume average). This average
can be either positive or negative but, if thermal, will converge to zero
at and above $T_c$.

Now, we are interested in determining whether the final field configuration
over large spatial scales is correlated to the configuration prior to 
reheating. Fig.~\ref{fig5} shows that only if one crosses $T_c$, by more than  
$+\hat \epsilon$, is the memory of the initial quenched configuration
erased (see in particular the two trajectories reaching higher temperatures
in comparison to the others). For these trajectories the field correlations
reach zero and after reheating evolve to a value manifestly different 
from that prior to reheating. 

For trajectories within the Ginzburg regime, that do not cross $T_c$, 
the change in the configuration of the order parameter as measured by 
Eq.~(\ref{cor}) is small.
In particular the field configuration existing before reheating is 
approximately recovered as the fields are cooled.  
The same is true for the string densities, including those of long strings.

Again we can understand these results using the tool developed 
in section \ref{FP}. The thermalization 
time, which is nothing else than the response time $\tau$ in the Kibble-Zurek
scenario, for long-wave length modes in the system is given by 
\begin{eqnarray}
t_{eq} \simeq {\eta \over m^2(T)}.
\end{eqnarray} 
We argued that it is a reasonable approximation to take $T$ to be the final 
temperature since the small scales in the system equilibrate much 
faster (provided  $m(T) < \eta $). 

Now $\hat t$, or equivalently $\hat \epsilon$, 
is defined as the time when the long wave-length modes in the
system can first respond to a change in bath temperature linear in time. 
It is computed by equating  the scaling of the response time $\tau(T)$
\begin{eqnarray}
\tau(\epsilon(T)) = {\eta \over m^2 \epsilon^{ \nu z}(T)}  
\end{eqnarray} 
to a linear change in time, imposed externally, i.e.
\begin{eqnarray}
\tau(\epsilon(\hat t)) = {\eta \over m^2 \epsilon^{\nu z} (\hat t)} = \hat t  
\end{eqnarray} 
$z$ is another critical 
exponent whose mean-field value is 2, see \cite{ABZ}.
This relation is usually solved by {\it assuming} a linear dependence in 
time for $\epsilon(t) = t /\tau_Q$, where $\tau_q$ is the rate of the 
external quench. 

Explicit calculation of $\Delta m^2(T)$ in (\ref{tadpole}) for our 
classical Boltzmann ensemble shows that 
\begin{eqnarray}
\Delta m^2(T) \simeq {\lambda \over \pi^2}  \Lambda T,
\label{tadpole2}
\end{eqnarray}
where $\Lambda$ is the ultraviolet cut off. This cutoff has physical 
meaning and is related to the breakdown of our scalar field model 
for high energy excitations, eg. fermionic quasi-particles in $^3He$. 
If indeed the external bath temperature is changed linearly then $T$ in 
Eq.~(\ref{tadpole2}) can be taken, over time scales larger 
than $\eta^{-1}$, to be linear in time. It then results trivially that 
$\epsilon(t)$ is also linear, which validates our assumption.

The  significance of $\hat t$ is that only when 
$\vert \epsilon(t) \vert  > \hat \epsilon$ can the long wave-length modes
in our system thermalize under an externally changing temperature
at a rate $\tau_Q$, i.e. keep pace with the externally imposed changes. 
Due to theoretical uncertainties the value 
of $\hat \epsilon$ adopted in Fig.~\ref{fig5}a was measured by monitoring 
the response of the system directly. Details are described elsewhere 
\cite{ABZ}.     

At the initial time, for temperature trajectories of Fig.~\ref{fig5},  
the system is in the process of breaking the $U(1)$ symmetry spontaneously, 
i.e. the expectation value of the $k=0$ mode of $\phi$, 
$\langle \phi \rangle$ is non-zero. 
Then, as the system is  heated towards $T_c$ equilibration of the long 
wave-length modes means that  
$\langle \phi \rangle \rightarrow 0$ and upon cooling show zero 
correlation in Fig.~\ref{fig5}b to its initial state. 
Since thermalization of $\langle \phi \rangle$ 
can only occur for $\epsilon \geq \hat \epsilon$, 
only the temperature trajectories crossing $+\hat \epsilon$ 
can attain zero correlations.   

It is then clear that the Ginzburg regime cannot change the symmetry breaking
process of the system,  including its associated long string configurations, 
unless a long amount of time is allowed. The Ginzburg regime 
is therefore less efficient at destroying topological defects 
(in the sense of requiring a longer time) than any other temperature
range outside the critical domain.

\section{Discussion and conclusions} 
\label{disc}

In this paper we have performed the most extensive analysis to date of the 
effects of large thermal fluctuations, within the Ginzburg regime, 
on the formation of topological defects. Our model field theory has 
already been studied extensively both in equilibrium and in tests 
of the theory of defect formation at temperature quenches, as
predicted by the critical dynamics of the theory. 

Under these controlled circumstances we analyzed critically the
assumptions underlying the traditional argument for the Ginzburg
temperature as the energy scale at which topological defects are formed.
We then proceeded to show that the effects of thermal fluctuations in
the Ginzburg regime upon a population of topological defects formed 
by the critical dynamics of the theory carries no particular signature
and leads mostly to small qualitative changes in the defect densities 
predicted by the theory of defect formation.

We have also shown that even prolonged exposure of a quenched 
field configuration to the Ginzburg  regime has little consequences in 
changing the order parameter configurations emerging at $-\hat \epsilon$, 
and associated string densities.
In addition we established that to truly destroy a quenched field 
configuration existing below $-\hat \epsilon$, one has to expose the system 
to temperatures well above $T_c$. In particular for a linear quench 
trajectory, a temperature of $ \epsilon \sim +\hat \epsilon$, must be reached 
in order to erase memory of the initial configuration.

These results were confirmed by analytical arguments based on the solutions 
of the associated Fokker-Plank equation. 
This analysis supports the conclusion that given the same amount of 
time of exposure to a thermal bath at a given temperature, the Ginzburg regime
is actually the least efficient range of temperatures at destroying 
the pattern of symmetry breaking inherited from criticallity. 
This includes topological defect configurations.

Our results fully support the theory of defect formation 
resulting from the critical dynamics of second order transitions 
\cite{Zurek} and all 
known thermodynamic results for vortex strings in $O(N)$ theories 
\cite{ABH,AB,Tsubo,XY}. In face of this
evidence we are lead to conclude that arguments singling out 
a special energy scale $T_G \neq T_c$, which would play an important 
role in defect formation rely on assumptions that are not realized in
the true (thermo)dynamics of our model and are thus invalid.

Thus we expect the results of this paper to carry over  
to the new Lancaster $^4He$ experiments. 
The results of reported in \cite{He4new} in these experiments cannot
therefore be attributed to the effects of Ginzburg regime in $^4He$.

\section*{Acknowledgments}
We thank  T.~Kibble, P.~Laguna and R.~Rivers 
for useful discussions. Numerical work 
was done on the T-division/CNLS Avalon Beowulf cluster, 
LANL. This research was supported by the U.S. Department 
of Energy, under contract W-7405-ENG-36.

\end{document}